\documentclass[sigconf,natbib=true,anonymous=false]{acmart}

\AtBeginDocument{%
  }

\setcopyright{acmlicensed}
\copyrightyear{2018}
\acmYear{2018}
\acmDOI{XXXXXXX.XXXXXXX}

\acmConference[Conference acronym 'XX]{Make sure to enter the correct
  conference title from your rights confirmation emai}{June 03--05,
  2018}{Woodstock, NY}
\acmISBN{978-1-4503-XXXX-X/18/06}

\usepackage{amsthm}
\usepackage{subcaption}
\usepackage{multirow} % for multi-column layout
\usepackage{hyperref}
\usepackage{tikz}
\usetikzlibrary{shapes.geometric}
\usepackage{cleveref}
\usepackage{graphicx}
\usepackage{pgf}
\usepackage{enumitem}
\usepackage{tikz}
\usetikzlibrary{positioning, arrows.meta, bending}
\usepackage[utf8]{inputenc}
\usepackage[T1]{fontenc}    % use 8-bit T1 fonts% 
\usepackage{pifont}
\usepackage{dsfont}
\usepackage{xcolor}
\usepackage{color}

\usepackage[normalem]{ulem}
\usepackage{float}
\usepackage{algorithm}
\usepackage{algorithmic}
\usepackage{lscape}
\usepackage{rotating}
\usepackage{longtable}

\usepackage{etoc}

\usepackage{mdframed}

\begin{document}

\title{Towards Robust Offline Evaluation: A Causal and Information Theoretic Framework for Debiasing Ranking Systems}

\author{Seyedeh Baharan Khatami}
\authornote{Both authors contributed equally to this research.}
\email{skhatami@ucsd.edu}
\affiliation{%
  \institution{Zillow Group}
  \city{Seattle}
  \state{WA}
  \country{USA}
  }
\orcid{1234-5678-9012}
\author{Sayan Chakraborty}
\authornotemark[1]
\email{sayanc@zillowgroup.com}
\affiliation{%
  \institution{Zillow Group}
  \city{Seattle}
  \state{WA}
  \country{USA}
  }
\author{Ruomeng Xu}
\email{ruomengx@zillowgroup.com}
\affiliation{%
  \institution{Zillow Group}
  \city{Seattle}
  \state{WA}
  \country{USA}
}

\author{Babak Salimi}
\email{bsalimi@ucsd.edu}
\affiliation{%
  \institution{UC San Diego}
  \city{San Diego}
  \state{CA}
  \country{USA}
}

\begin{abstract}
Evaluating retrieval-ranking systems is crucial for developing high-performing models. While online A/B testing is the gold standard, its high cost and risks to user experience require effective offline methods. However, relying on historical interaction data introduces biases—such as selection, exposure, conformity, and position biases—that distort evaluation metrics, driven by the Missing-Not-At-Random (MNAR) nature of user interactions and favoring popular or frequently exposed items over true user preferences.

We propose a novel framework for robust offline evaluation of retrieval-ranking systems, transforming MNAR data into Missing-At-Random (MAR) through reweighting combined with black-box optimization, guided by neural estimation of information-theoretic metrics. Our contributions include (1) a causal formulation for addressing offline evaluation biases, (2) a system-agnostic debiasing framework, and (3) empirical validation of its effectiveness. This framework enables more accurate, fair, and generalizable evaluations, enhancing model assessment before deployment.
\end{abstract}


\maketitle

\section{INTRODUCTION}
Recommender systems (RS) help users navigate information overload by providing personalized suggestions, benefiting both users and providers. Companies refine RS models through iterative improvements, underscoring the need for robust evaluation. While A/B testing remains the gold standard, it is costly, slow, and risks degrading user experience. Offline evaluation using historical data offers a more efficient alternative but suffers from biases due to its observational nature, including selection \citep{yang2018unbiased, surveybias}, exposure \citep{surveybias}, popularity \citep{chen2020esam, 10.1145/3269206.3269264, surveybias}, and position biases \citep{collins2018study, surveybias}. Implicit feedback only captures positive interactions, leading to Missing-Not-At-Random (MNAR) data, where engagement skews toward frequently surfaced items, making user preferences harder to infer. Ignoring these biases in offline evaluation can reinforce the long-tail effect, propagate biases from prior models, and misalign results with A/B tests, increasing the risk of poor model selection \citep{10.1145/1639714.1639717, schnabel2016recommendations, 10.1145/1835804.1835895}.

Existing debiasing approaches have limitations: some assume Missing-At-Random (MAR) interactions, which is unrealistic \citep{he2016vbpr, 10.1145/3038912.3052639, 10.1145/2792838.2799671}; others address single biases like position \citep{10.1145/3437963.3441794} or popularity bias \citep{10.1145/3269206.3269264} but lack generalizability. Many methods require clean, unbiased data, which is often impractical \citep{hu2008collaborative, 10.1145/3404835.3463118}.

To bridge these gaps, our framework allows for specifying a bias attribute—such as exposure, popularity, or temporal bias—and debiases the evaluation data accordingly by transforming MNAR data into Missing-At-Random (MAR) data, ensuring a more reliable assessment of recommendation quality. Unlike methods that require access to a clean, bias-free dataset, our approach operates effectively without such data but can leverage it when available for further debiasing. Additionally, our framework is generalizable, system-agnostic, and adaptable across diverse ranking systems, making it suitable for a wide range of recommendation scenarios. By leveraging a causal formulation and an information-theoretic perspective, our method corrects for biases inherent in evaluation data, leading to offline metrics that more accurately reflect true user preferences and system performance. Our contributions are threefold:
\begin{itemize}
    \item \textbf{Causal Problem Formulation:} A theoretical foundation leveraging information-theoretic principles to mitigate biases in offline evaluations.
    \item \textbf{Mutual Information-Based Framework:} A general system-agnostic approach that minimizes dependence between observed interactions and a given biasing factor.
    \item \textbf{Empirical Validation:} Evaluation on both public and company internal offsite real-time recommendation system data, demonstrating effectiveness as a minimal-effort debiasing prerequisite for ranking systems.
\end{itemize}

\section{RELATED WORK}
Evaluating retrieval-ranking systems while mitigating biases in historical interaction data has been extensively studied. Inverse Propensity Scoring (IPS) \cite{yang2018unbiased, 10.1145/1639714.1639717, 10.1145/3404835.3463118, hu2008collaborative} is a widely used method for correcting selection bias but suffers from high variance. To improve stability, doubly robust estimators \cite{pmlr-v97-wang19n}. Adversarial learning techniques \cite{xu2020adversarial, wu2021debiasgan} aim to identify and mitigate bias through adversarial training, while causal inference methods \cite{Wang2021DeconfoundedRF, zhang2022causal, wang2023counterfactual, Zhang2021CausalIF} address confounding factors by leveraging do-calculus and backdoor adjustment.

Recent works explore invariant learning to disentangle user preferences from bias \cite{Zhang2021CausalIF, zheng2022cbr}, but these approaches struggle with accuracy and stability. Knowledge distillation methods \cite{Bai2024InvariantDL} fuse invariant and variant information to improve generalization. Adaptive model selection strategies \cite{10.1145/3706637} dynamically switch between biased and debiased models depending on test conditions. 
Despite these advancements, most methods require unbiased data supervision, which is costly and challenging in real-world settings. Our proposed framework addresses this limitation by introducing a resampling-based evaluation method using conditional mutual information to systematically mitigate the entanglement between bias and user preferences, offering a scalable and generalizable solution.
\section{PROBLEM FORMULATION}
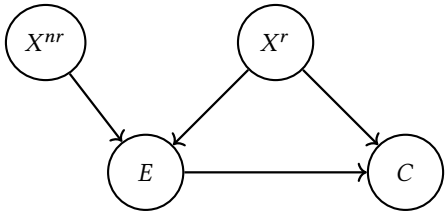
\begin{figure}[t]
\centering
% \begin{subfigure}[t]{0.45\textwidth}
% \begin{minipage}{.5\textwidth}
% \centering
\begin{tikzpicture}[node distance=2cm and 2cm, scale=0.1]
  % \node[draw, ellipse, fill=white, minimum size=1cm, thick] (S) {\large $S$};
  % \node[align=left, below=0.1cm of S] (S_desc){};

    \node[draw, ellipse, fill=white, minimum size=1cm, thick] (Xp) {\large $X^{nr}$};
  \node[align=left, below=0.1cm of Xp] (Xp_desc){};

  \node[draw, ellipse, fill=white, right= 2cm of Xp, minimum size=1cm, thick] (X) {\large $X^r$};
  \node[align=left, right=0.1cm of X] (X_desc) {};

  \node[draw, ellipse, fill=white, below left=1cm and 1cm of X, minimum size=1cm, thick] (E) {\large $E$};
  \node[align=left, below=0.1cm of E] (E_desc){};

  \node[draw, ellipse, fill=white, below right=1cm and 1cm of X, minimum size=1cm, thick] (C) {\large $C$};
  \node[align=left, below=0.1cm of C] (C_desc) {};
  
  \draw[->, line width=0.3mm] (Xp) -- (E);
  \draw[->, line width=0.3mm] (X) -- (E);
  \draw[->, line width=0.3mm] (X) -- (C);
  \draw[->, line width=0.3mm] (E) -- (C);
\end{tikzpicture}
\caption{Causal DAG of Random Variables: \(X^r\) represents relevant user-specific item features, \(X^{nr}\) represents non-relevant biasing factor feature influencing exposure, \(E\) is exposure, and \(C\) is click.}
% \end{subfigure}
\label{fig:DAG}
\vspace{-3mm}
\end{figure}

This section introduces the notation and theoretical formulation of the problem. Our dataset is defined as 
\(\mathcal{D} = \{(U_i, I_i, X_i, E_i, C_i) \; | \; i = 1, 2, \dots, N \}\), 
where \(U_i\) and \(I_i\) are user and item IDs for the \(i\)-th interaction. 
The feature vector \(X_i = (X^r_i, X^{nr}_i)\) consists of user-specific item features relevant to the user's preferences \(X^r_i\) and exposure-related features \(X^{nr}_i\), which may not reflect user preferences. 
\(E_i\) denotes whether \(I_i\) was exposed to \(U_i\), and \(C_i\) is the observed interaction (e.g., a click). The goal is to reduce the impact of $X^{nr}$ on exposure, as it biases interactions ($C$).  
We aim to debias with respect to $X^{nr}$ and illustrate this with four examples, though the approach extends to other exposure-related biases as well.

\begin{itemize}
    % \item \textbf{Systematic Confounding Bias:}  
    % Let $X^{nr}$ represent the relevance scores assigned by the ranking system. Given ranking systems has the ability to expose the items to a users and simultaneously collect engagement signals based on the exposures to run iterations on top of the current system, it can create it's own confirmation bias in the collected data where the system itself works as a confounder when analyzing the engagement signals.

    \item \textbf{Popularity Bias:}  
    Let $X^{nr}$ represent item popularity scores. High popularity may skew exposure, overexposing few items and underexposing others, limiting visibility.

    \item \textbf{Sensitive Attribute Bias:}  
    Let $X^{nr}$ represent a sensitive attribute (e.g., gender or race) that biases exposure, leading to unequal visibility of items associated with certain groups.

    \item \textbf{Staleness Bias:}  
    Let $X^{nr}$ denote the timestamp an item was introduced. Systems may favor items added earlier, giving them higher exposure probabilities. This staleness bias is a key case study in our experiments, discussed in detail in the evaluation section.
    
\end{itemize}

The definition of $X^{nr}$ varies by application and is typically determined using domain knowledge from system designers.

We observe samples from the joint distribution $P(X, C)$, denoted as the observed distribution $P_{\mathbf{o}}(X, C)$. If $X^{nr}$ can be partitioned into $m$ groups $\{X_1^{nr}, X_2^{nr}, \dots, X_m^{nr}\}$, an ideal ranking system should satisfy $P_{\mathbf{o}}(X^{r}, C \mid X_k^{nr}) = P_{\mathbf{o}}(X^{r}, C \mid X_l^{nr})$ for all $k, l \in \{1, 2, \dots, m\}$. However, this equality often fails due to various systematic factors or design choices in ranking systems. In general, a biased ranking system generates varying signals from different segments of $X^{nr}$, leading to discrepancies in exposure across these segments. This disproportionate exposure reduces the likelihood of user interactions in certain areas. The resulting Missing-Not-At-Random (MNAR) data in implicit feedback datasets biases the evaluation, limiting its ability to detect performance shifts triggered by changes in regions with insufficient exposure. Consequently, evaluation results may not accurately reflect the true performance of new models.

In the true distribution $P(X, C)$, which reflects users' preferences, user-item interactions should be independent of system exposure $E$ (an observed proxy of $X^{nr}$), given the user-specific item features relevant to the user's preferences. This is because users' intrinsic preferences are assumed to be unaffected by the system’s exposure choices. Formally, this implies the conditional independence $C \perp E \mid X$. However, in observed data, this independence is violated, as exposure ($E$) directly biases interactions ($C$), often due to irrelevant factors in $X^{nr}$.

\section{METHOD}

In this section, we introduce a general debiasing framework designed to address the specified bias attribute, as formulated in the previous section.

\subsection{GENERAL FRAMEWORK}
\begin{figure}[t]
  \centering
  \includegraphics[height=0.25\textheight, width=\columnwidth]{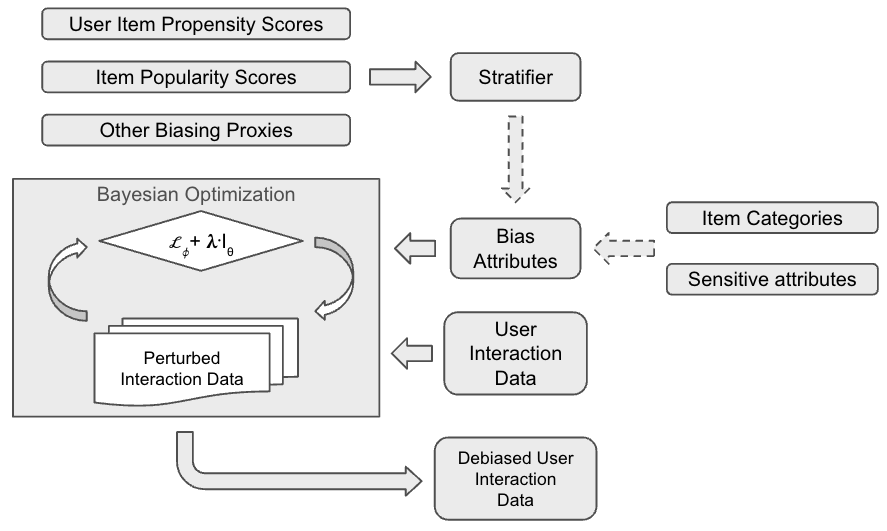}
  \caption{Framework schema: Continuous bias attributes are bucketized into user-defined bins. The bias attribute is passed to the Bayesian optimization framework, which optimizes the objective function—comprising CMI estimation and click prediction performance—to find the optimal resampling weights, defined over the bias attribute bins.}
  \label{fig:framework}
  \vspace{-3mm}
\end{figure}

 The proposed debiasing framework employs a conditional independence guided process for data perturbation (Figure \ref{fig:framework}). The bias attribute, $X^{nr}$, is defined based on the system’s use case and expert input, representing factors affecting exposure mechanisms, item popularity, or user-item interactions like propensity scores. If clean data is available, our framework can enhance debiasing by incorporating it into biased data and applying the proposed approach. However, our method remains effective even without it.

The framework perturbs biased data by sampling rows based on bias attributes to minimize the conditional dependence between $C$ and $E$ given $X^r$. Depending on the problem, this dependence can be measured at the exposure or bias attribute layer in the causal graph. For instance, popularity bias needs to be tracked not only through item popularity measures but also by assessing how the system interacts with or mitigates these effects, making exposure-layer measurement preferable. Conversely, for debiasing item ratings with suspected self-selection bias, proxies like propensity scores can be directly used. Our approach treats the exposure variable as a mediator, capturing various bias attributes while ensuring adaptability to different recommender system biases. The bias attribute is explicitly used in optimization to ensure debiasing with respect to $X^{nr}$, measuring conditional independence between $E$ and $C$ given $X^r$, though a similar framework could replace $E$ with specific attributes $X^{nr}$ when feasible.

The perturbation begins by defining $K$ weights, $w_1, \dots, w_K$, based on the bias attribute. For continuous $X^{nr}$, the attribute is discretized into $K$ bins (as shown in Figure \ref{fig:framework}), where $K$ is a user-specified parameter controlling stratification granularity. For categorical $X^{nr}$, $K$ corresponds to the number of classes. Each interaction data row is assigned to a group based on its bias attribute value. The perturber then resamples the data with probabilities proportional to the assigned group weights. The resampled perturbed data is then passed to the conditional dependence estimator to assess dependence between $E$ and $C$ given $X^r$. Besides ensuring conditional independence, perturbations should also preserve click prediction utility. Therefore, we optimize a utility metric alongside the conditional dependence measure to maintain the joint distribution structure of $(X, C)$.

Since the loss function depends on resampling weights and is not differentiable, we employ black-box optimization techniques to determine the optimal weights. The next subsection details the objective function and black-box optimization approach.

\subsection{FRAMEWORK SPECIFICATION}
We use conditional mutual information (CMI) to measure the conditional dependence between users' true preferences and the bias attribute, formulated as follows:  
{\small \begin{equation}
\label{eq1}
    I(E; C \mid X^r) = \mathbb{E}_{X^r}\left[\sum_{E, C} P(E, C \mid X^r) \log \frac{P(E, C \mid X^r)}{P(E \mid X^r) P(C \mid X^r)}\right] 
\end{equation}}

As noted earlier, the formulation can use $X^{nr}$ instead of $E$, depending on the problem. Estimating CMI directly through conditional density estimation with finite data can be challenging and lead to biased results. To address this, we use the dual representation of KL-divergence, known as the Donsker-Varadhan representation, as shown in \cite{belghazi2018mutual}, formulated as:

{\small \begin{equation}
\label{eq2}
I(E; C \mid X)) = \sup_{T: \mathcal{E} \times \mathcal{C} \mid \mathcal{X} \rightarrow \mathbb{R}}\mathbb{E}_{P_{EC\mid X}}[T(X)] - log(\mathbb{E}_{P_{E\mid X}\bigotimes P_{C\mid X}}[e^{T(X)}])
\end{equation}}

where $T$ is restricted to be the family of functions $T_{\theta}: \mathcal{E} \times \mathcal{C} \mid \mathcal{X} \rightarrow \mathbb{R}$ parametrized by a neural network with parameters $\theta \in \Theta$. The objective can be maximized by gradient ascent.

To preserve the structure of the joint distribution of $X^r$ and $C$ and maintain the predictive power of the ranking system via $X^r$, we train a separate model $f_{\phi}: \mathcal{X} \rightarrow \mathcal{C}$, with the following loss:

{\small
\begin{equation}
\label{eqBCE}
\mathcal{L}_{\phi} = -\frac{1}{N} \sum_{} \left[ c_i \log(f_{\phi(x^r_i)}) + (1 - c_i) \log(1 - f_{\phi(x^r_i)}) \right]
\end{equation}
}

The joint loss is then defined as:

\begin{equation}
\label{eq3}
\mathcal{L} = \mathcal{L_{\phi}} + \lambda\cdot I_{\theta}(E; C \mid X^r))
\end{equation}
where $\lambda$ balances the prediction loss and the
regularization. Specifically, we utilize Bayesian Optimization \citep{inbook, snoek2012practical} as our black-box optimization framework to minimize the proposed loss and optimize the sampling weights for perturbing the data. The pseudo-code for our algorithm is provided in Algorithm \ref{alg:debiasing} and Figure \ref{fig:framework} illustrates the framework components.

% due to biased representation of the feature space X which could highly distort the conditional distributions referred in equation (\ref{eq1}). 
{\small
\begin{algorithm}
\caption{Guided CMI based Debiasing of RS Data}
\label{alg:debiasing}
\begin{algorithmic}[1] % The "[1]" adds line numbering
\REQUIRE $\mathcal{D} = (U, I, X^r, X^{nr}, E, C)$, number of bins $K$, number of iterations $n_{iter}$, $\lambda$
\ENSURE Debiased dataset $\mathcal{D}_{\text{debiased}}$
\STATE N = size of $\mathcal{D}$
\STATE Discretize $X^{nr}$ into $K$ bins if $X^{nr}$ is continuous
% \STATE $optimizer = LL+CMI(w_1, \dots, w_K)$

\WHILE{$n_{iter}$}
\STATE $\mathcal{D}^\prime \leftarrow sample(\mathcal{D}, [w_1, \dots, w_K])$
\STATE $\theta \leftarrow \text{Train}(T_{\theta})$
\STATE $\phi \leftarrow \text{Train}(f_{\phi})$
\STATE $\mathcal{L} = \mathcal{L_{\phi}} + \lambda\cdot I_{\theta}(E; C \mid X^r)$
\STATE $[w_1, \dots, w_K] = optimizer.minimize(\mathcal{L})$
\ENDWHILE

\STATE $\mathcal{D}_{\text{debiased}} \leftarrow sample(\mathcal{D}, [{w_1}_{opt}, \dots, {w_K}_{opt}])$
\RETURN $\mathcal{D}_{\text{debiased}}$
\end{algorithmic}
\end{algorithm}}

\vspace{-3mm}
\section{EXPERIMENTS}
This section details the empirical evaluation of our framework via public data and company internal offsite real-time recommendation system data.

\subsection{Coat Data}

To evaluate our method's performance, we use the Coat dataset \cite{schnabel2016recommendations}, designed for selection bias evaluation in recommendation systems. It consists of $290$ users, $300$ coats, $6960$ MNAR training ratings, and $4640$ MAR test ratings. The explicit $1$-$5$ ratings enable a controlled comparison between biased and unbiased performance.

As our goal is to generate reliable, unbiased ranking evaluation data, we assess the generated data, debiased using different mechanisms, against MAR golden data. Ideally, $C$ should be conditionally independent of $X^{nr}$ given $X^{r}$, i.e., $P(C \mid X^{r}) = P(C \mid X^{r}, X^{nr})$, where we use $X^{nr}$ instead of $E$ to frame the selection bias problem. 

Following \cite{schnabel2016recommendations}, we estimate Naïve Bayes propensity scores to categorize bias attributes into five strata. We use CatBoost for click prediction, training and evaluating it under scenarios in Table \ref{tab:table1}. The target is binarized as ratings $\geq 4$ or below. The debiasing process is model-agnostic, allowing substitution of CatBoost with any model.

\begin{table}[t]
  \label{table1}
  \centering
  \resizebox{\columnwidth}{!}{%
  \begin{tabular}{|c|c|c|c|c|}
    \hline
    Methods & AUC & Precision & Recall & F1 \\
    \hline
    E1: BT + BenchE & 0.791 & 0.664 & 0.221 & 0.332\\
    E2: BT + BiasE & 0.751 (-5.1\%) & 0.454 (-31.6\%) & 0.698 (+215.8\%) & 0.555 (+67.2\%) \\
    E3: BT + DBiasIPSE & 0.760 (-3.9\%) & 0.655 (-1.4\%) & 0.202 (-8.6\%) & 0.308 (-7.2\%) \\
    E4: BT + StratEval & 0.760 (-3.9\%) & 0.683 (+2.9\%) & 0.234 (+5.9\%) & 0.349 (+5.1\%) \\
    E5: BT + DBiasCMIE & 0.772 (-2.4\%) & 0.654 (-1.5\%) & 0.239 (+7.5\%) & 0.349 (+5.1\%) \\
    \hline
    T1: BT (\textbackslash w BF) + BenchE & 0.792 & 0.686 & 0.231 & 0.346 \\
    T2: IPS-Train + BenchE & 0.789  & 0.687 & 0.227 & 0.341 \\
    T3: DBiasCMIT + BenchE & 0.788 (-0.03\%) & 0.658 (-4.2\%) & 0.235 (+3.5\%) & 0.346 (+1.5\%) \\
    T4: DBiasCMIT (\textbackslash w BF) + BenchE & 0.790 & 0.675 & 0.217 & 0.329 \\
    \hline
  \end{tabular}
  }
  \caption{  Performance of various perturbation mechanisms on training and evaluation sets. The top half ($E$ prefix) evaluates debiasing the evaluation data, while the bottom half ($T$ prefix) focuses on debiasing the training data. BT: Biased Training; BenchE: Benchmark Evaluation; BiasE: Biased Evaluation; DBiasCMIE: CMI Debiased Evaluation; DBiasCMIT: CMI Debiased Training; DBiasIPSE: IPS Debiased Evaluation; StratEval: Propensity Stratified Evaluation; \textbackslash w BF: bias factor is included in training.}
    \vspace{-8mm}
  
  \label{tab:table1}
\end{table} 

We analyze our method's performance from two perspectives in Table \ref{tab:table1}. The top half evaluates models trained on biased data and tested on debiased datasets generated by different methods. The ideal scenario, $E1$, uses MAR golden data as the benchmark, with other methods compared by relative fluctuation. We perturb 10\% of the evaluation set and assess whether our debiased evaluation serves as a reliable proxy for a randomized benchmark. We compare our CMI-based debiasing method to biased evaluation data ($E2$), IPS-based debiasing \cite{yang2018unbiased} ($E3$), and stratified evaluation \cite{jadidinejad2021simpson} ($E4$). Evaluation on biased data ($E2$) performs poorly, showing a large gap from the unbiased benchmark ($E1$), highlighting the need for debiasing. Our method ($E5$) shows the lowest drift from $E1$ in AUC and $F_1$-score, which balances recall and precision, compared to all baselines.

 Debiasing can also be applied to training data to improve model performance by better capturing users' true preferences. The bottom half of Table \ref{tab:table1} compares our method ($T3$: perturbing 10\% of training data) with IPS-based debiasing ($T2$) and training on biased data ($E1$), all evaluated on randomized benchmark data. The bottom half of the table shows that our CMI-based debiasing method ($T3$) improves recall and F1-score, with a slight precision drop and marginal AUC decrease. These changes indicate bias reduction, as the precision drop alongside increased recall and F1 suggests better capture of true user preferences over selection bias.

To assess debiasing's impact on click prediction, we introduce the biasing feature (propensity scores) in $E1$ and $T3$ (resulting in $T1$ and $T4$ models) to evaluate $P(C \mid X^{r}, X^{nr})$. $T1$ outperforms $E1$ across all metrics, showing that including $X^{nr}$ in pre-perturbation distributions aids prediction. In post-perturbation distributions, the gap between $T4$ and $T3$ narrows, with $T3$ achieving higher recall and F1, indicating that our perturbations effectively reduce bias dependence.

\subsection{Internal Offsite Recommendation Data}

To assess the real-world impact of our method, we use internal user interaction data from an offsite recommendation system exhibiting staleness bias. The system limits daily notifications to avoid overwhelming users, resetting the count each morning. Since recommendations rely on external events and user relevance, earlier events post-reset have a higher chance of being sent.

To address this bias, we implement debiasing by stratifying interactions into hourly buckets. We incorporate onsite recommendation data, free from staleness bias, to enhance the debiasing process. Our goal is to debias user clicks and compare performance against a more reliable down-funnel signal—saves. Unlike clicks, saves occur through multiple channels (onsite and offsite) and are less biased, but their sparsity makes them unsuitable for training or evaluation, highlighting the need to debias the more prevalent click data.

\begin{figure}[t]
  \centering
  \includegraphics[height=0.25\textheight, width=\columnwidth]{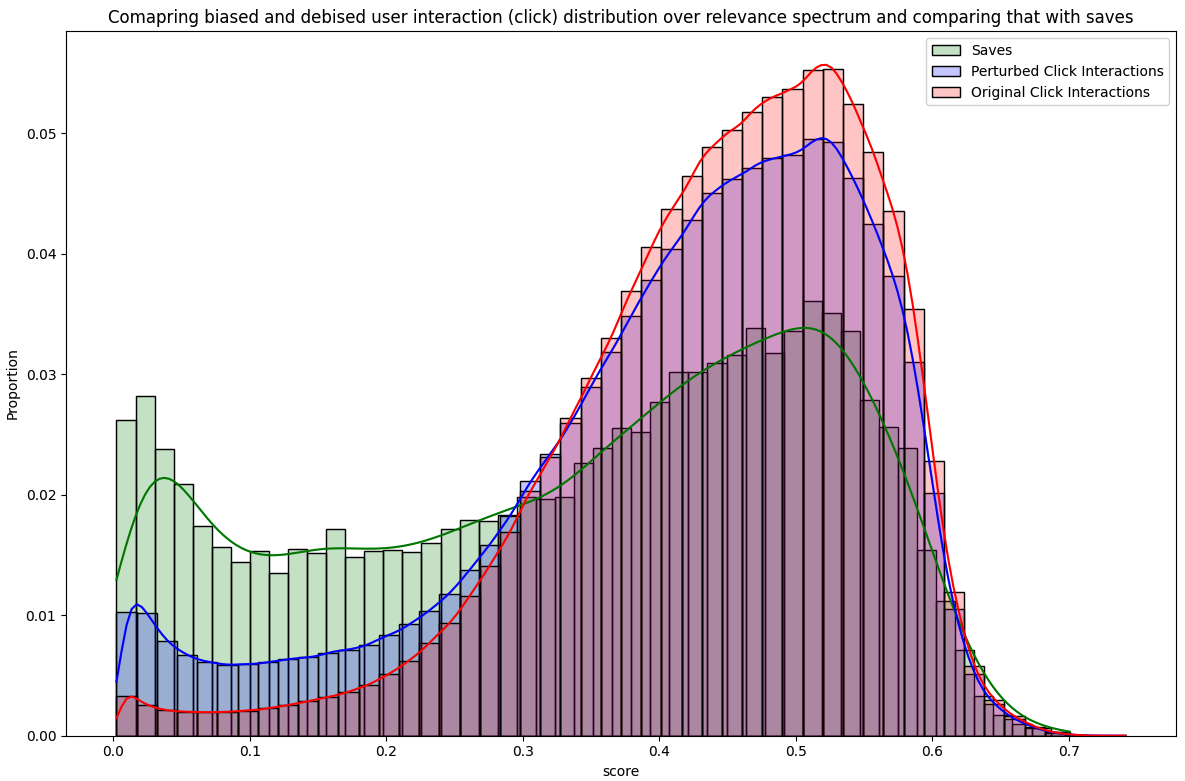}
  \caption{Comparing the impact of debiasing user interactions across relevance spectrum vs. down-funnel preference signals (e.g., saves)}
  \label{fig:debias}
\end{figure}

Figure \ref{fig:debias} shows the distribution of user interactions before and after debiasing across the relevance spectrum. The system creates a feedback loop, where early events receive higher relevance scores. The debiasing rebalances the click distribution to align with the save distribution, making upper-funnel interactions better reflect true user preferences, enabling more reliable training and evaluation of ranking systems.

Our goal is to minimize the gap between $P(C\mid X^{r})$ and $P(C\mid X^{r}, X^{nr})$ to enhance the conditional independence of $C$ and $X^{nr}$. To quantify the distributional shift, we use the Wasserstein Distance between the two density functions. We model these densities by training two CatBoost models—one excluding and one including the biasing factor in the feature space. As shown in Table \ref{tab:table2}, the Wasserstein Distance decreases after perturbation, indicating a weaker dependency on $X^{nr}$. Additionally, the reduced density gap between $X_{(c)}^{r}$ and $X_{(s)}^{r}$ further validates our observations in Figure \ref{fig:debias}, where \(X_{(c)}^{r}\) and \(X_{(s)}^{r}\) represent subsets of the feature space where users clicked and saved items, respectively, after exposures based on \(P(C \mid X^{r})\).

\begin{table}[t]
  \label{table2}
  \centering
  \resizebox{\columnwidth}{!}{%
  \begin{tabular}{|c|c|c|}
    \hline
    Data & $W(P(C\mid X^{r})$, $P(C\mid X^{r}, X^{nr}))$ & $W(P(C\mid X^{r}\in X_{(c)}^{r}), P(C\mid X^{r}\in X_{(s)}^{r}))$ \\
    \hline
    Original Data & 0.043 & 0.099\\
    Perturbed Data &  0.038 & 0.070\\
    \hline
  \end{tabular}
  }
  \caption{Wasserstein Distances ($W$). $X_{(c)}^{r}$ and $X_{(s)}^{r}$ represent subsets of the feature space where users clicked and saved items, respectively, after exposures based on $P(C \mid X^{r})$.}
  \label{tab:table2}
  \vspace{-4mm}
\end{table}

\section{CONCLUSION}
We propose a model-agnostic framework to mitigate biases in recommender system evaluation using a causal perspective applicable to various bias attributes. By addressing the impact of non-relevant features on user interactions, our approach perturbs the data to reduce dependence between exposure and interaction, conditioned on relevant features, via neural estimation of conditional mutual information optimized with Bayesian Optimization. Our framework uses row-level perturbation and black-box optimization, presenting scalability challenges. Future work will explore counterfactual data augmentation at the feature level to enhance efficiency and scalability.

\bibliographystyle{ACM-Reference-Format}
\bibliography{references}


\begin{thebibliography}{28}

%%% the Web address.

\ifx \showCODEN    \undefined \def \showCODEN     #1{\unskip}     \fi
\ifx \showDOI      \undefined \def \showDOI       #1{#1}\fi
\ifx \showISBNx    \undefined \def \showISBNx     #1{\unskip}     \fi
\ifx \showISBNxiii \undefined \def \showISBNxiii  #1{\unskip}     \fi
\ifx \showISSN     \undefined \def \showISSN      #1{\unskip}     \fi
\ifx \showLCCN     \undefined \def \showLCCN      #1{\unskip}     \fi
\ifx \shownote     \undefined \def \shownote      #1{#1}          \fi
\ifx \showarticletitle \undefined \def \showarticletitle #1{#1}   \fi
\ifx \showURL      \undefined \def \showURL       {\relax}        \fi
% invisible to TeX
\providecommand\bibfield[2]{#2}
\providecommand\bibinfo[2]{#2}
\providecommand\natexlab[1]{#1}
\providecommand\showeprint[2][]{arXiv:#2}

\bibitem[Bai et~al\mbox{.}(2024)]%
        {Bai2024InvariantDL}
\bibfield{author}{\bibinfo{person}{Ting Bai}, \bibinfo{person}{Weijie Chen}, \bibinfo{person}{Cheng Yang}, {and} \bibinfo{person}{Chuan Shi}.} \bibinfo{year}{2024}\natexlab{}.
\newblock \showarticletitle{Invariant debiasing learning for recommendation via biased imputation}.
\newblock \bibinfo{journal}{\emph{Inf. Process. Manag.}}  \bibinfo{volume}{62} (\bibinfo{year}{2024}), \bibinfo{pages}{104028}.
\newblock
\urldef\tempurl%
\url{https://api.semanticscholar.org/CorpusID:274850767}
\showURL{%
\tempurl}


\bibitem[Belghazi et~al\mbox{.}(2018)]%
        {belghazi2018mutual}
\bibfield{author}{\bibinfo{person}{Mohamed~Ishmael Belghazi}, \bibinfo{person}{Aristide Baratin}, \bibinfo{person}{Sai Rajeshwar}, \bibinfo{person}{Sherjil Ozair}, \bibinfo{person}{Yoshua Bengio}, \bibinfo{person}{Aaron Courville}, {and} \bibinfo{person}{Devon Hjelm}.} \bibinfo{year}{2018}\natexlab{}.
\newblock \showarticletitle{Mutual information neural estimation}. In \bibinfo{booktitle}{\emph{International conference on machine learning}}. PMLR, \bibinfo{pages}{531--540}.
\newblock


\bibitem[Chen et~al\mbox{.}(2023)]%
        {surveybias}
\bibfield{author}{\bibinfo{person}{Jiawei Chen}, \bibinfo{person}{Hande Dong}, \bibinfo{person}{Xiang Wang}, \bibinfo{person}{Fuli Feng}, \bibinfo{person}{Meng Wang}, {and} \bibinfo{person}{Xiangnan He}.} \bibinfo{year}{2023}\natexlab{}.
\newblock \showarticletitle{Bias and Debias in Recommender System: A Survey and Future Directions}.
\newblock \bibinfo{journal}{\emph{ACM Trans. Inf. Syst.}} \bibinfo{volume}{41}, \bibinfo{number}{3}, Article \bibinfo{articleno}{67} (\bibinfo{date}{Feb.} \bibinfo{year}{2023}), \bibinfo{numpages}{39}~pages.
\newblock
\showISSN{1046-8188}
\urldef\tempurl%
\url{https://doi.org/10.1145/3564284}
\showDOI{\tempurl}


\bibitem[Chen et~al\mbox{.}(2020)]%
        {chen2020esam}
\bibfield{author}{\bibinfo{person}{Zhihong Chen}, \bibinfo{person}{Rong Xiao}, \bibinfo{person}{Chenliang Li}, \bibinfo{person}{Gangfeng Ye}, \bibinfo{person}{Haochuan Sun}, {and} \bibinfo{person}{Hongbo Deng}.} \bibinfo{year}{2020}\natexlab{}.
\newblock \showarticletitle{Esam: Discriminative domain adaptation with non-displayed items to improve long-tail performance}. In \bibinfo{booktitle}{\emph{Proceedings of the 43rd International ACM SIGIR Conference on Research and Development in Information Retrieval}}. \bibinfo{pages}{579--588}.
\newblock


\bibitem[Collins et~al\mbox{.}(2018)]%
        {collins2018study}
\bibfield{author}{\bibinfo{person}{Andrew Collins}, \bibinfo{person}{Dominika Tkaczyk}, \bibinfo{person}{Akiko Aizawa}, {and} \bibinfo{person}{Joeran Beel}.} \bibinfo{year}{2018}\natexlab{}.
\newblock \showarticletitle{A study of position bias in digital library recommender systems}.
\newblock \bibinfo{journal}{\emph{arXiv preprint arXiv:1802.06565}} (\bibinfo{year}{2018}).
\newblock


\bibitem[He and McAuley(2016)]%
        {he2016vbpr}
\bibfield{author}{\bibinfo{person}{Ruining He} {and} \bibinfo{person}{Julian McAuley}.} \bibinfo{year}{2016}\natexlab{}.
\newblock \showarticletitle{VBPR: visual bayesian personalized ranking from implicit feedback}. In \bibinfo{booktitle}{\emph{Proceedings of the AAAI conference on artificial intelligence}}, Vol.~\bibinfo{volume}{30}.
\newblock


\bibitem[Hsieh et~al\mbox{.}(2017)]%
        {10.1145/3038912.3052639}
\bibfield{author}{\bibinfo{person}{Cheng-Kang Hsieh}, \bibinfo{person}{Longqi Yang}, \bibinfo{person}{Yin Cui}, \bibinfo{person}{Tsung-Yi Lin}, \bibinfo{person}{Serge Belongie}, {and} \bibinfo{person}{Deborah Estrin}.} \bibinfo{year}{2017}\natexlab{}.
\newblock \showarticletitle{Collaborative Metric Learning}. In \bibinfo{booktitle}{\emph{Proceedings of the 26th International Conference on World Wide Web}} (Perth, Australia) \emph{(\bibinfo{series}{WWW '17})}. \bibinfo{publisher}{International World Wide Web Conferences Steering Committee}, \bibinfo{address}{Republic and Canton of Geneva, CHE}, \bibinfo{pages}{193–201}.
\newblock
\showISBNx{9781450349130}
\urldef\tempurl%
\url{https://doi.org/10.1145/3038912.3052639}
\showDOI{\tempurl}


\bibitem[Hu et~al\mbox{.}(2008)]%
        {hu2008collaborative}
\bibfield{author}{\bibinfo{person}{Yifan Hu}, \bibinfo{person}{Yehuda Koren}, {and} \bibinfo{person}{Chris Volinsky}.} \bibinfo{year}{2008}\natexlab{}.
\newblock \showarticletitle{Collaborative filtering for implicit feedback datasets}. In \bibinfo{booktitle}{\emph{2008 Eighth IEEE international conference on data mining}}. Ieee, \bibinfo{pages}{263--272}.
\newblock


\bibitem[Jadidinejad et~al\mbox{.}(2021)]%
        {jadidinejad2021simpson}
\bibfield{author}{\bibinfo{person}{Amir~H Jadidinejad}, \bibinfo{person}{Craig Macdonald}, {and} \bibinfo{person}{Iadh Ounis}.} \bibinfo{year}{2021}\natexlab{}.
\newblock \showarticletitle{The simpson’s paradox in the offline evaluation of recommendation systems}.
\newblock \bibinfo{journal}{\emph{ACM Transactions on Information Systems (TOIS)}} \bibinfo{volume}{40}, \bibinfo{number}{1} (\bibinfo{year}{2021}), \bibinfo{pages}{1--22}.
\newblock


\bibitem[Krishnan et~al\mbox{.}(2018)]%
        {10.1145/3269206.3269264}
\bibfield{author}{\bibinfo{person}{Adit Krishnan}, \bibinfo{person}{Ashish Sharma}, \bibinfo{person}{Aravind Sankar}, {and} \bibinfo{person}{Hari Sundaram}.} \bibinfo{year}{2018}\natexlab{}.
\newblock \showarticletitle{An Adversarial Approach to Improve Long-Tail Performance in Neural Collaborative Filtering}. In \bibinfo{booktitle}{\emph{Proceedings of the 27th ACM International Conference on Information and Knowledge Management}} (Torino, Italy) \emph{(\bibinfo{series}{CIKM '18})}. \bibinfo{publisher}{Association for Computing Machinery}, \bibinfo{address}{New York, NY, USA}, \bibinfo{pages}{1491–1494}.
\newblock
\showISBNx{9781450360142}
\urldef\tempurl%
\url{https://doi.org/10.1145/3269206.3269264}
\showDOI{\tempurl}


\bibitem[Lee et~al\mbox{.}(2021)]%
        {10.1145/3404835.3463118}
\bibfield{author}{\bibinfo{person}{Jae-woong Lee}, \bibinfo{person}{Seongmin Park}, {and} \bibinfo{person}{Jongwuk Lee}.} \bibinfo{year}{2021}\natexlab{}.
\newblock \showarticletitle{Dual Unbiased Recommender Learning for Implicit Feedback}. In \bibinfo{booktitle}{\emph{Proceedings of the 44th International ACM SIGIR Conference on Research and Development in Information Retrieval}} (Virtual Event, Canada) \emph{(\bibinfo{series}{SIGIR '21})}. \bibinfo{publisher}{Association for Computing Machinery}, \bibinfo{address}{New York, NY, USA}, \bibinfo{pages}{1647–1651}.
\newblock
\showISBNx{9781450380379}
\urldef\tempurl%
\url{https://doi.org/10.1145/3404835.3463118}
\showDOI{\tempurl}


\bibitem[Lim et~al\mbox{.}(2015)]%
        {10.1145/2792838.2799671}
\bibfield{author}{\bibinfo{person}{Daryl Lim}, \bibinfo{person}{Julian McAuley}, {and} \bibinfo{person}{Gert Lanckriet}.} \bibinfo{year}{2015}\natexlab{}.
\newblock \showarticletitle{Top-N Recommendation with Missing Implicit Feedback}. In \bibinfo{booktitle}{\emph{Proceedings of the 9th ACM Conference on Recommender Systems}} (Vienna, Austria) \emph{(\bibinfo{series}{RecSys '15})}. \bibinfo{publisher}{Association for Computing Machinery}, \bibinfo{address}{New York, NY, USA}, \bibinfo{pages}{309–312}.
\newblock
\showISBNx{9781450336925}
\urldef\tempurl%
\url{https://doi.org/10.1145/2792838.2799671}
\showDOI{\tempurl}


\bibitem[Marlin and Zemel(2009)]%
        {10.1145/1639714.1639717}
\bibfield{author}{\bibinfo{person}{Benjamin~M. Marlin} {and} \bibinfo{person}{Richard~S. Zemel}.} \bibinfo{year}{2009}\natexlab{}.
\newblock \showarticletitle{Collaborative prediction and ranking with non-random missing data}. In \bibinfo{booktitle}{\emph{Proceedings of the Third ACM Conference on Recommender Systems}} (New York, New York, USA) \emph{(\bibinfo{series}{RecSys '09})}. \bibinfo{publisher}{Association for Computing Machinery}, \bibinfo{address}{New York, NY, USA}, \bibinfo{pages}{5–12}.
\newblock
\showISBNx{9781605584355}
\urldef\tempurl%
\url{https://doi.org/10.1145/1639714.1639717}
\showDOI{\tempurl}


\bibitem[Mockus et~al\mbox{.}(2014)]%
        {inbook}
\bibfield{author}{\bibinfo{person}{J. Mockus}, \bibinfo{person}{Vytautas Tiesis}, {and} \bibinfo{person}{Antanas Zilinskas}.} \bibinfo{year}{2014}\natexlab{}.
\newblock \bibinfo{booktitle}{\emph{The application of Bayesian methods for seeking the extremum}}. Vol.~\bibinfo{volume}{2}.
\newblock \bibinfo{pages}{117--129}.
\newblock
\showISBNx{0-444-85171-2}


\bibitem[Oosterhuis and de~Rijke(2021)]%
        {10.1145/3437963.3441794}
\bibfield{author}{\bibinfo{person}{Harrie Oosterhuis} {and} \bibinfo{person}{Maarten de Rijke}.} \bibinfo{year}{2021}\natexlab{}.
\newblock \showarticletitle{Unifying Online and Counterfactual Learning to Rank: A Novel Counterfactual Estimator that Effectively Utilizes Online Interventions}. In \bibinfo{booktitle}{\emph{Proceedings of the 14th ACM International Conference on Web Search and Data Mining}} (Virtual Event, Israel) \emph{(\bibinfo{series}{WSDM '21})}. \bibinfo{publisher}{Association for Computing Machinery}, \bibinfo{address}{New York, NY, USA}, \bibinfo{pages}{463–471}.
\newblock
\showISBNx{9781450382977}
\urldef\tempurl%
\url{https://doi.org/10.1145/3437963.3441794}
\showDOI{\tempurl}


\bibitem[Schnabel et~al\mbox{.}(2016)]%
        {schnabel2016recommendations}
\bibfield{author}{\bibinfo{person}{Tobias Schnabel}, \bibinfo{person}{Adith Swaminathan}, \bibinfo{person}{Ashudeep Singh}, \bibinfo{person}{Navin Chandak}, {and} \bibinfo{person}{Thorsten Joachims}.} \bibinfo{year}{2016}\natexlab{}.
\newblock \showarticletitle{Recommendations as treatments: Debiasing learning and evaluation}. In \bibinfo{booktitle}{\emph{international conference on machine learning}}. PMLR, \bibinfo{pages}{1670--1679}.
\newblock


\bibitem[Snoek et~al\mbox{.}(2012)]%
        {snoek2012practical}
\bibfield{author}{\bibinfo{person}{Jasper Snoek}, \bibinfo{person}{Hugo Larochelle}, {and} \bibinfo{person}{Ryan~P Adams}.} \bibinfo{year}{2012}\natexlab{}.
\newblock \showarticletitle{Practical Bayesian optimization of machine learning algorithms}. In \bibinfo{booktitle}{\emph{Advances in Neural Information Processing Systems}}, Vol.~\bibinfo{volume}{25}. \bibinfo{pages}{2951--2959}.
\newblock


\bibitem[Steck(2010)]%
        {10.1145/1835804.1835895}
\bibfield{author}{\bibinfo{person}{Harald Steck}.} \bibinfo{year}{2010}\natexlab{}.
\newblock \showarticletitle{Training and testing of recommender systems on data missing not at random}. In \bibinfo{booktitle}{\emph{Proceedings of the 16th ACM SIGKDD International Conference on Knowledge Discovery and Data Mining}} (Washington, DC, USA) \emph{(\bibinfo{series}{KDD '10})}. \bibinfo{publisher}{Association for Computing Machinery}, \bibinfo{address}{New York, NY, USA}, \bibinfo{pages}{713–722}.
\newblock
\showISBNx{9781450300551}
\urldef\tempurl%
\url{https://doi.org/10.1145/1835804.1835895}
\showDOI{\tempurl}


\bibitem[Wang et~al\mbox{.}(2021)]%
        {Wang2021DeconfoundedRF}
\bibfield{author}{\bibinfo{person}{Wenjie Wang}, \bibinfo{person}{Fuli Feng}, \bibinfo{person}{Xiangnan He}, \bibinfo{person}{Xiang Wang}, {and} \bibinfo{person}{Tat seng Chua}.} \bibinfo{year}{2021}\natexlab{}.
\newblock \showarticletitle{Deconfounded Recommendation for Alleviating Bias Amplification}.
\newblock \bibinfo{journal}{\emph{Proceedings of the 27th ACM SIGKDD Conference on Knowledge Discovery \& Data Mining}} (\bibinfo{year}{2021}).
\newblock
\urldef\tempurl%
\url{https://api.semanticscholar.org/CorpusID:235166201}
\showURL{%
\tempurl}


\bibitem[Wang et~al\mbox{.}(2019)]%
        {pmlr-v97-wang19n}
\bibfield{author}{\bibinfo{person}{Xiaojie Wang}, \bibinfo{person}{Rui Zhang}, \bibinfo{person}{Yu Sun}, {and} \bibinfo{person}{Jianzhong Qi}.} \bibinfo{year}{2019}\natexlab{}.
\newblock \showarticletitle{Doubly Robust Joint Learning for Recommendation on Data Missing Not at Random}. In \bibinfo{booktitle}{\emph{Proceedings of the 36th International Conference on Machine Learning}} \emph{(\bibinfo{series}{Proceedings of Machine Learning Research}, Vol.~\bibinfo{volume}{97})}, \bibfield{editor}{\bibinfo{person}{Kamalika Chaudhuri} {and} \bibinfo{person}{Ruslan Salakhutdinov}} (Eds.). \bibinfo{publisher}{PMLR}, \bibinfo{pages}{6638--6647}.
\newblock
\urldef\tempurl%
\url{https://proceedings.mlr.press/v97/wang19n.html}
\showURL{%
\tempurl}


\bibitem[Wang et~al\mbox{.}(2023)]%
        {wang2023counterfactual}
\bibfield{author}{\bibinfo{person}{Y. Wang}, \bibinfo{person}{J. Li}, \bibinfo{person}{J. Wu}, {and} \bibinfo{person}{X. Liu}.} \bibinfo{year}{2023}\natexlab{}.
\newblock \showarticletitle{Counterfactual Learning for Debiasing User Representations in Recommender Systems}. In \bibinfo{booktitle}{\emph{Proceedings of the 36th AAAI Conference on Artificial Intelligence}}. \bibinfo{pages}{11245--11252}.
\newblock
\urldef\tempurl%
\url{https://doi.org/10.1609/aaai.v36i10.21471}
\showDOI{\tempurl}


\bibitem[Wang et~al\mbox{.}(2025)]%
        {10.1145/3706637}
\bibfield{author}{\bibinfo{person}{Zimu Wang}, \bibinfo{person}{Hao Zou}, \bibinfo{person}{Jiashuo Liu}, \bibinfo{person}{Jiayun Wu}, \bibinfo{person}{Pengfei Tian}, \bibinfo{person}{Yue He}, {and} \bibinfo{person}{Peng Cui}.} \bibinfo{year}{2025}\natexlab{}.
\newblock \showarticletitle{AdaptSel: Adaptive Selection of Biased and Debiased Recommendation Models for Varying Test Environments}.
\newblock \bibinfo{journal}{\emph{ACM Trans. Knowl. Discov. Data}} \bibinfo{volume}{19}, \bibinfo{number}{2}, Article \bibinfo{articleno}{29} (\bibinfo{date}{Jan.} \bibinfo{year}{2025}), \bibinfo{numpages}{39}~pages.
\newblock
\showISSN{1556-4681}
\urldef\tempurl%
\url{https://doi.org/10.1145/3706637}
\showDOI{\tempurl}


\bibitem[Wu et~al\mbox{.}(2021)]%
        {wu2021debiasgan}
\bibfield{author}{\bibinfo{person}{Chuhan Wu}, \bibinfo{person}{Fangzhao Wu}, {and} \bibinfo{person}{Yongfeng Huang}.} \bibinfo{year}{2021}\natexlab{}.
\newblock \showarticletitle{DEBIASGAN: eliminating position bias in news recommendation with adversarial learning}.
\newblock \bibinfo{journal}{\emph{arXiv preprint arXiv:2106.06258}} (\bibinfo{year}{2021}).
\newblock


\bibitem[Xu et~al\mbox{.}(2020)]%
        {xu2020adversarial}
\bibfield{author}{\bibinfo{person}{Da Xu}, \bibinfo{person}{Chuanwei Ruan}, \bibinfo{person}{Evren Korpeoglu}, \bibinfo{person}{Sushant Kumar}, {and} \bibinfo{person}{Kannan Achan}.} \bibinfo{year}{2020}\natexlab{}.
\newblock \showarticletitle{Adversarial counterfactual learning and evaluation for recommender system}.
\newblock \bibinfo{journal}{\emph{Advances in Neural Information Processing Systems}}  \bibinfo{volume}{33} (\bibinfo{year}{2020}), \bibinfo{pages}{13515--13526}.
\newblock


\bibitem[Yang et~al\mbox{.}(2018)]%
        {yang2018unbiased}
\bibfield{author}{\bibinfo{person}{Longqi Yang}, \bibinfo{person}{Yin Cui}, \bibinfo{person}{Yuan Xuan}, \bibinfo{person}{Chenyang Wang}, \bibinfo{person}{Serge Belongie}, {and} \bibinfo{person}{Deborah Estrin}.} \bibinfo{year}{2018}\natexlab{}.
\newblock \showarticletitle{Unbiased offline recommender evaluation for missing-not-at-random implicit feedback}. In \bibinfo{booktitle}{\emph{Proceedings of the 12th ACM conference on recommender systems}}. \bibinfo{pages}{279--287}.
\newblock


\bibitem[Zhang et~al\mbox{.}(2022)]%
        {zhang2022causal}
\bibfield{author}{\bibinfo{person}{J. Zhang}, \bibinfo{person}{X. Zhao}, \bibinfo{person}{Y. Zhang}, {and} \bibinfo{person}{H. Liu}.} \bibinfo{year}{2022}\natexlab{}.
\newblock \showarticletitle{Causal Intervention for Mitigating Popularity Bias in Recommendation}. In \bibinfo{booktitle}{\emph{Proceedings of the 28th ACM SIGKDD Conference on Knowledge Discovery and Data Mining}}. \bibinfo{pages}{925--933}.
\newblock
\urldef\tempurl%
\url{https://doi.org/10.1145/3534678.3539110}
\showDOI{\tempurl}


\bibitem[Zhang et~al\mbox{.}(2021)]%
        {Zhang2021CausalIF}
\bibfield{author}{\bibinfo{person}{Yang Zhang}, \bibinfo{person}{Fuli Feng}, \bibinfo{person}{Xiangnan He}, \bibinfo{person}{Tianxin Wei}, \bibinfo{person}{Chonggang Song}, \bibinfo{person}{Guohui Ling}, {and} \bibinfo{person}{Yongdong Zhang}.} \bibinfo{year}{2021}\natexlab{}.
\newblock \showarticletitle{Causal Intervention for Leveraging Popularity Bias in Recommendation}.
\newblock \bibinfo{journal}{\emph{Proceedings of the 44th International ACM SIGIR Conference on Research and Development in Information Retrieval}} (\bibinfo{year}{2021}).
\newblock
\urldef\tempurl%
\url{https://api.semanticscholar.org/CorpusID:234482660}
\showURL{%
\tempurl}


\bibitem[Zheng et~al\mbox{.}(2022)]%
        {zheng2022cbr}
\bibfield{author}{\bibinfo{person}{Zhi Zheng}, \bibinfo{person}{Zhaopeng Qiu}, \bibinfo{person}{Tong Xu}, \bibinfo{person}{Xian Wu}, \bibinfo{person}{Xiangyu Zhao}, \bibinfo{person}{Enhong Chen}, {and} \bibinfo{person}{Hui Xiong}.} \bibinfo{year}{2022}\natexlab{}.
\newblock \showarticletitle{CBR: context bias aware recommendation for debiasing user modeling and click prediction}. In \bibinfo{booktitle}{\emph{Proceedings of the ACM Web Conference 2022}}. \bibinfo{pages}{2268--2276}.
\newblock


\end{thebibliography}

\appendix

\end{document}